\documentclass[aps,prd,preprint,showpacs]{revtex4}
\usepackage{graphicx}
\begin{document}
\draft
\title{Effects of the CP Odd Dipole Operators on Gluino Production at Hadron Colliders}
\author{A.T. Alan}
\thanks{e-mail: alan\_a@ibu.edu.tr}
\address{Department of Physics, Abant Izzet Baysal University, 14280, Bolu, Turkey}
\date{\today}
\pacs{12.38.Lg, 12.60.Jv, 14.80.Ly}
\begin{abstract}

We present the cross sections for the hadroproduction of gluinos
by taking into account the CP odd dipole operators in
supersymmetric QCD. The dependence of the cross sections on these
operators is analyzed for the hadron colliders the Tevatron
($\sqrt S$=1.8 TeV) and the Cern LHC ($\sqrt{S}$=14 TeV). The
enhancement of the hadronic cross section is obviously mass
dependent and for a 500 GeV gluino, is up to 16 \% (over 73 pb) at
the LHC while it is 8 \% (over 0.63 fb) at the Tevatron.

\end{abstract}
\maketitle

\section{Introduction}

Supersymmetric QCD (SUSY-QCD) is based on the colored particles of
the Minimal Supersymmetric Standard Model (MSSM) \cite{HG,HP}
spectrum; quarks, gluons and their superpartners squarks and
gluinos. Since supersymmetry is a broken one rather than an exact
symmetry, the masses of the superparticles extremely exceed the
masses of their SM partners \cite{LMT}. Upper bound limits of
$\mathcal{O}$(1 TeV) are set to these masses for the sake of the
solution of the hierarchy problem. In most of the analysis the
scalar partners of the five light quarks are assumed mass
degenerate \cite{WRMZ}.

Searching for supersymmetric particles will be one of the main
goals of the future experimental program of high energy physics.
The particles in the strong interaction sector can be searched for
most efficiently at hadron colliders. As they are presently
searched at the Tevatron ($\sqrt s$= 1.8 TeV) the Large Hadron
Collider (LHC) with center of mass energy of $\sqrt s$= 14 TeV
will in a sense be a gluino factory.

At the fundamental level SUSY receives some additional effects
from the existent particles predicted by high energy models so
called GUT or String Theory. As these effect are quite general we
consider them in gluino pair production. At this level we know the
interaction vertices in supersymmetric QCD. After the prediction
of Kane and Leville \cite{KL} for the tree level hadronic
production of gluinos several improvements for this process have
been performed \cite{WRMZ}. The production of gluino pairs in
electron-positron annihilation is analyzed in ref 4. In the
present analysis we reconsider the productions of gluino pairs in
hadron-hadron collisions and generalize to include the CP odd
terms to investigate the effects of these CP violating operators
in these processes. We assumed an updated range of 300-500 GeV for
the gluino masses.

\section{Gluino Production with the CP Violating Terms}

The dominant contributions to the production cross sections of
gluino pairs in $pp $ or $p\bar{p}$ collisions come from the
subprocesses  $q\bar q \rightarrow \tilde{g}\tilde{g}$ and $g g
\rightarrow \tilde{g}\tilde{g}$. The relevant Feynman diagrams are
displayed in Figures 1 and 2 and the differential cross sections
are calculated in the Appendix. In the first subprocess
$q\bar{q}\rightarrow \tilde g\tilde g$, in principle $\hat t$ and
$\hat u $-channel squark exchanges have also contributions in
addition to the annihilation s channel gluion exchange. But these
contributions are almost always negligible since squarks of first
and second generations must be nearly degenerate and they are
heavy to satisfy the Electric Dipole Moment bounds. Therefore we
will consider only the s-channel via gluon exchange for the
reaction $q\bar q \rightarrow \tilde{g}\tilde{g}$. In calculations
of the differential cross sections we use the standard structure
of the effective CP-odd lagrangian including the Weinberg operator
and color EDMs of quarks \cite{DOKM,DMA,AP} and the interaction
lagrangian of the gluinos with the gauge field gluons \cite{HG} to
obtain the following three interaction vertices;

i) Quark-quark-gluon
\begin{figure}[h]
  \includegraphics[width=10cm]{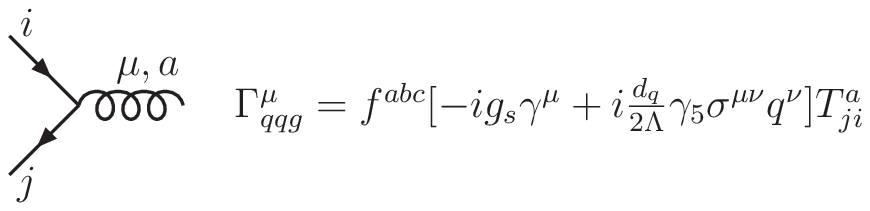}\\
\end{figure}

ii) Gluon-gluino-gluino
\begin{figure}[h]
  \includegraphics[width=10cm]{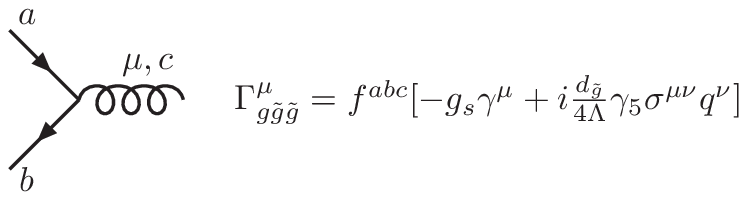}\\
\end{figure}

iii) Three gluon
\begin{figure}[h]
  \includegraphics[width=14cm]{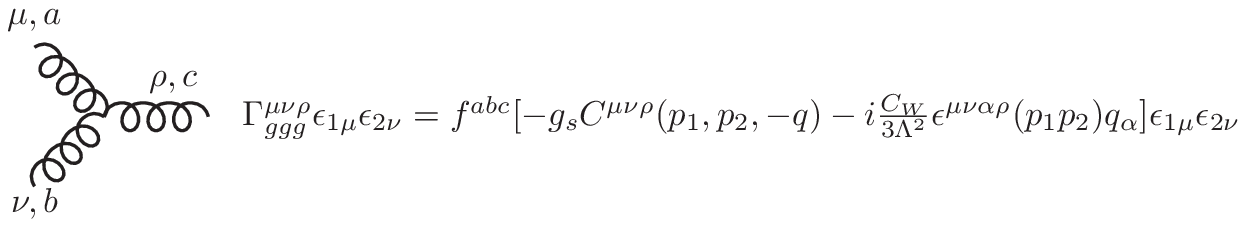}\\
\end{figure}

where $g_s$ is the strong coupling constant, $q$ is the momentum
carried by the mediator, $d_q$ and $\tilde d_g$ are effective
color EDMs couplings of quarks and gluinos respectively. $\Lambda$
denotes the scale up to which the effective theory is assumed to
hold. $C^{\mu\nu\rho}(p_1,p_2,q)$ is the standard three gluon
vertex with  all momenta incoming
$C^{\mu\nu\rho}(p_1,p_2,q)=g^{\mu\nu}(p_1-p_2)^{\rho}+g^{\nu\rho}(p_2-q)^{\mu}+g^{\mu\rho}(q-p_1)^{\nu}
$. $C_W$ is the coefficient of the Weinberg operator and the term
with this coefficient is the result of a straightforward
calculation of the dimension-6 Weinberg operator \cite{SW} :
\begin{eqnarray}
\mathcal{O}_W=-\frac{C_W}{6}f_{abc}\epsilon^{\mu\lambda\nu\rho}G_a^{\mu\lambda}G^b_{\nu\sigma}G^c_{\rho\sigma}
\end{eqnarray}
where $G_a^{\mu\nu}$ is gluon field strength tensor.

The differential cross sections are presented in the Appendix.
Integrating them over $\hat t$ leads to the total partonic cross
sections:

\begin{eqnarray}
 \hat{\sigma}_q (q_i \bar q_i\rightarrow \tilde{g}\tilde{g})&=& ( \frac{1}{72\pi \hat s^4}) [J( d^2{d_1}^2J^2\hat s^2
 - g_s^2\hat s( ( d^2 + {d_1}^2 )J^2 - 12d^2m^2\hat s \nonumber \\
&& + 3( d^2 + {d_1}^2 ) \hat s^2 )
      + g_s^4( J^2 + 3\hat s( 4m^2 + \hat s)
        )  ) ]
\end{eqnarray}
where $\hat s =x_1 x_2 S $ is the partonic CM energy,  $J=s\beta$,
$\beta=\sqrt{1-\frac{4m^2}{\hat s}}$ being gluino velocity in the
partonic center of mass system, $m$ denotes mass of the produced
gluinos, and $d_1=d_q/2\Lambda$, $d=d_{\tilde g}/4\Lambda$. In
obtaining Eq.1 only the s-channel of Fig.1 was considered.

For the second  subprocess $gg \rightarrow \tilde{g}\tilde{g}$
contributions of all the three diagrams in Fig.2 are considered,
and the integrated cross section in this case is obtained as

\begin{eqnarray}
\hat{\sigma }(gg \rightarrow
\tilde{g}\tilde{g})&=&\frac{1}{2}\frac{1}{512\pi \hat s^4( J^2 -
\hat s^2)}[3( J(d^4\hat s^2( -5J^4 + 6912m^8 + 3354m^4\hat s^2 -
480m^2\hat s^3 + 9\hat s^4 \nonumber\\
&&+ J^2( -3354m^4 + 480m^2\hat s - 4\hat s^2 ))- 4g_s^4(J^4 +  2J^2\hat s( -12m^2 + 7\hat s)\nonumber \\
&& - 3\hat s^2( 64m^4 - 8m^2\hat s + 5\hat s^2))- 2d^2g_s^2\hat s( J^4 + J^2( 144m^4 + 210m^2\hat s - 25\hat s^2)\nonumber\\
&&+ 6\hat s( 384m^6 - 56m^4\hat s - 35m^2\hat s^2 + 4\hat s^3))\nonumber\\
&&+ 12d m( 4g_s^2 +d^2( 20m^2 - 3\hat s))\hat s^3(\hat s^2 -J^2)w\\
&&+ 12\hat s^3( J^2 - \hat s^2)( d^2\hat s( -4m^2 + \hat s) + 2g_s^2( 2m^2 + \hat s))w^2)\nonumber\\
&&- 12\hat s(\hat s^2 - J^2)( 4g_s^4( -4m^4 + 2m^2\hat s + \hat s^2)\nonumber\\
&& + d^3m^4( d( 48m^4
+ 316m^2\hat s + 3\hat s^2) + 16m\hat s^2w)\nonumber\\
&& + 2dg_s^2m( 4d m( 12m^4 - 12m^2\hat s + \hat s^2) + \hat
s^3w))\log (\frac{J + \hat s}{J-\hat s}))]\nonumber
\end{eqnarray}
with and $w=C_W/6\Lambda^2$. Since the gluinos are Majarona
fermions the statistical factor of 1/2 is implied to avoid double
counting.

The total cross section $pp (p\bar p)\rightarrow g\tilde{g} X$ is
calculated as a function of the hadronic CM energy $S$ and gluino
mass $m$, by convoluting the cross sections of subprocesses and
parton densities through the factorization theorem

\begin{eqnarray}
\sigma_{tot} (s,m)=  \int_{\tau_0}^1d\tau \int_\tau^1
\frac{dx}{x}\frac{1}{1+\delta_{ij}} \sum_{i j}\left[f_i
(x)f_j(\frac{\tau}{x})+f_i(\frac{\tau}{x})f_j(x)\right]\hat\sigma(m,\hat
s,\hat u)
\end{eqnarray}
where $(i,j)=(g,g)$, $(q_{\alpha},\bar q_{\alpha})$,
$\alpha=u,d,s$. In performing the calculations we used the Feynman
gauge for the internal gluon propagators. We have also used
$-g^{\mu\nu}$ for the polarization sum of the external gluons by
adding the ghost term. In Tables I-IV we have presented the cross
sections for the gluino masses of 300, 400 and 500 GeV to see the
$\Lambda$ and mass dependence. The calculated cross sections are
not very sensitive to the $\Lambda$ values but drop fast with the
increasing mass values. The $gg$ initial state is always dominant
at the LHC. At the Tevatron for m=300 GeV the $q\bar q$ cross
sections are grater than the $gg$ cases, but the situation is
reversed for the masses of 400 or 500 GeV.

\begin{table}
  \centering
\begin{tabular}{c|ccc}
  \hline\hline
  $\Lambda$ (GeV) & m=300 GeV & m=400 GeV  & m=500 GeV  \\
  \hline
  800& 0.227689 & 0.0136852& 0.000536575 \\
  1000&0.221508&0.0131209&0.000506254\\
  2000& 0.2133299 & 0.0123716 & 0.000465985 \\
  4000& 0.211253& 0.0121849& 0.000455946 \\
  6000& 0.210874 & 0.0121503 & 0.000454088 \\
  8000& 0.210741& 0.0121382 & 0.000453438\\
\hline\hline

\end{tabular}
\caption{The hadronic cross sections in pb for the $q\bar q$
initial states at the Tevatron ($\sqrt s$= 1.8 TeV)}
\end{table}

\begin{table}
  \centering
\begin{tabular}{c|ccc}
  \hline\hline
  $\Lambda$ (GeV) & m=300 (GeV) & m=400 (GeV)  & m=500 (GeV)  \\
  \hline
  800 & 0.352658  & 0.00868101    & 0.000174379   \\
  1000& 0.352641& 0.0086626 & 0.000172692\\
  2000& 0.353933& 0.00873083  & 0.000174725  \\
  4000& 0.354495& 0.0087646  & 0.000176001  \\
  6000& 0.354610& 0.00877161  & 0.000176271  \\
  8000& 0.354652& 0.00877412  & 0.000176369  \\
\hline\hline

\end{tabular}
\caption{The hadronic cross sections in pb for the $gg$ initial
states at the Tevatron ($\sqrt s$= 1.8 TeV)}
\end{table}

\begin{table}
\centering
\begin{tabular}{c|ccc}
  \hline\hline
  $\Lambda$ (GeV) & m=300 (GeV) & m=400 (GeV)  & m=500 (GeV)  \\
  \hline
  800& 25.9082 & 12.8225& 7.37316 \\
  1000&24.1935&11.5765&6.43782\\
  2000& 22.0388 & 10.0294 &5.28865 \\
  4000& 21.5236& 9.66296& 5.01886 \\
  6000& 21.4292 & 9.59601 &4.96966 \\
  8000& 21.3963 & 9.57264 &4.9525 \\
\hline\hline
\end{tabular}
\caption{The hadronic cross sections in pb for the $q\bar q$
initial states at the LHC ($\sqrt s$= 14 TeV)}
\end{table}

\begin{table}
\centering
\begin{tabular}{c|ccc}
  \hline\hline
  $\Lambda$ (GeV) & m=300 (GeV) & m=400 (GeV)  & m=500 (GeV)  \\
  \hline
  800& 324.288 & 163.190& 91.125 \\
 1000& 306.743& 148.258&78.410 \\
  2000& 293.053& 137.326& 69.599 \\
  4000& 291.167 & 136.018 &68.681 \\
  6000& 290.884 & 135.837 &68.565 \\
  8000&290.791& 135.778&68.529\\
\hline\hline
\end{tabular}
\caption{The hadronic cross sections in pb for the $gg$ initial
states at the LHC ($\sqrt s$= 14 TeV)}
\end{table}
\section{Discussion and Conclusion}

The results for the $p\bar p(pp)\rightarrow \tilde g \tilde g X$
cross sections are presented in Fig.3 for $q\bar q$ annihilation
and in Fig.4 for $gg$ fusion at the Tevatron by fixing
$\Lambda$=1000 GeV. Similarly the corresponding results are
presented in Fig.5 and Fig.6 at the LHC. The solid lines are
displayed for $d$=$d_1$=$w$=0, without CP violating terms and
dotted lines are displayed by taking the contributions of CP-odd
terms. For illustrations all effective couplings $d_q$, $d_{\tilde
g}$ and $C_W$ are set equal to 1. We used MRST \cite{MRST}
parametrization in convoluting of parton densities. We treated the
gluon and light quark flavors as massless. From Fig.4 it is
obvious that effects of the CP violating terms in lagrangian are
negligible at the Tevatron for the $gg$ fusion. The enhancements
in the total hadronic cross sections at this collider for gluinos
of masses 300, 400, and 500 GeV are 1.5 \%, 4.2 \%, and 7.9 \%
respectively. However the event rates are very low due to the low
cross sections; for instance the number of events for 500 GeV
gluinos at the Tevatron with an integrated luminosity of 2
$fb^{-1}$ is only 1-2 per year.

At the LHC, the event rates are substantially high; for instance
the number of events for 500 GeV gluinos is as high as $10^7$ in
each LHC detector for a high integrated luminosity of 100
$fb^{-1}$.

In addition to the high event rates at the LHC, CP odd terms give
extra contributions, for instance the enhancements in the total
hadronic cross sections for 300, 400, and 500 GeV gluinos are 6
\%, 10 \%, and 16 \% respectively. Main contribution to the
enhancements comes from the Weinberg term, especially at high mass
values. Finally gluon-gluon fusion is always the dominant process
in pp collisions such that the LHC can be treated as a gluon-gluon
collider, at a first approximation.

\begin{acknowledgements}
I am grateful to D.A. Demir for many valuable discussions. This
work was supported in part by the Abant Izzet Baysal University
Research Found.
\end{acknowledgements}
\appendix
\section{Color Factors}
The generators of SU$_c$(3) are defined by the commutation
relations
\begin{eqnarray}
[T_a,T_b]=if_{abc}T_c
\end{eqnarray}
The matrices defined as $F_{bc}^a=-if_{abc}$ also satisfy similar
relations $[F_a,F_b]=if_{abc}F_c$ which means that $F_a$ form the
adjoint representation of SU$_c$(3). Fundamental identities we
have used in this work for the traces of color matrices in the
fundamental and adjoint representations are
\begin{eqnarray}
&&Tr T_a=0\nonumber \\
&&Tr(T^aT^b)=\frac{1}{2}\delta_{ab}\nonumber\\
&&Tr(F_aF_b)=3\delta_{ab}\nonumber\\
&&Tr(F_aF_bF_c)=i\frac{3}{2}f_{abc}\nonumber\\
&&Tr(F_aF_bF_aF_c)=\frac{9}{2}\delta_{bc}\nonumber\\
&&f_{acd}f_{bcd}=3\delta_{ab}\nonumber\\
&&f_{abc}f_{abc}=24
\end{eqnarray}

Because of different color factors, it is convenient to express
the square of the matrix element in the form
\begin{eqnarray}
|M(gg\rightarrow \tilde{g}\tilde
g)|^2&=&C_{ss}|M_s|^2+C_{tt}|M_t|^2+C_{uu}|M_u|^2+2C_{st}M_s\bar{M}_t-2C_{su}M_s\bar{M}_u\\
&&-2C_{tu}M_t\bar{M}_u\nonumber
\end{eqnarray}
where the obtained values of color factors are

\begin{eqnarray}
C_{ss}=C_{tt}=C_{uu}=72\nonumber\\
C_{st}=C_{su}=-C_{tu}=36
\end{eqnarray}
The color factor of $|M(q\bar q\rightarrow \tilde{g}\tilde g)|^2$
is obtained to be 12.

\section{Differential Cross Sections}
The differential cross sections to produce gluino pairs are
determined from the Feynman diagrams of Figures 1 and 2 via the
subprocesses $q\bar q\rightarrow\tilde g\tilde g$ and
$gg\rightarrow\tilde g\tilde g$ respectively and obtained as

\begin{eqnarray}
\frac{d\hat{\sigma }(q\bar q \rightarrow
\tilde{g}\tilde{g})}{d\hat{t}}&=&\frac{1}{12\,\pi \,\hat
s^4}[d^2{d_1}^2{( \hat t -
\hat u ) }^2{( -2m^2 + \hat t + \hat u )}^2\nonumber \\
&& + 2g_s^4( 6m^4 + \hat t^2 + \hat u^2 - 4m^2( \hat t + \hat u ))\\
&& - 4g_s^2( 2m^2 - \hat t - \hat u )(({d_1}^2( m^2 - \hat t)( m^2
- \hat u )) + d^2( m^4 - \hat t\hat u))]\nonumber
\end{eqnarray}

by considering only s-channel of Fig.1, and

\begin{eqnarray}
\frac{d\hat{\sigma }(gg \rightarrow
\tilde{g}\tilde{g})}{d\hat{t}}&=&\frac{9}{512\pi\hat
s^2}[M_{s}^2+M_{t}^2+M_u^2+M_{st}+M_{su}+M_{tu}]
\end{eqnarray}
with

\begin{eqnarray}
M_s^2&=&\frac{4}{\hat s^2}[ 4\,g_s^4( m^2 - \hat t) ( m^2 - \hat
u) - w^2d^2\,{( -2\,m^2 + \hat t + \hat u ) }^4( 2\,m^2 + \hat t + \hat u)\nonumber\\
&&+ g_s^2( 2\,m^2 - \hat t - \hat u )( d^2\,{( \hat t - \hat
u)}^2+ 2\,w^2( 4\,m^2 - \hat t - \hat u ) \,{( -2\,m^2 + \hat t +
\hat
u ) }^2)]\nonumber\\\nonumber\\
M_t^2&=&\frac{-2}{(\hat t-m^2)^2}[4\,g_s^4( m^4 - \hat t\,\hat u +
m^2( 3\,\hat t + \hat u))+ 2\,g_s^2\,d^2( 5\,m^6 -
59\,m^4\,\hat t + 3\,\hat t^3 + m^2\,\hat t( 7\,\hat t - 4\,\hat u))\nonumber\\
&& + d^4\,\hat t( 4\,m^6 + 67\,m^4\,\hat t + \hat t^2( 4\,\hat t -
\hat
u) + m^2\,\hat t( 69\,\hat t + \hat u))]\nonumber\\\nonumber\\
M_u^2&=&\frac{-2}{(\hat u-m^2)^2}[ 4\,g_s^4( m^4 - \hat t\,\hat u
+ m^2( \hat t + 3\,\hat u)) + 2\,g_s^2\,d^2( 5\,m^6 -
59\,m^4\,\hat u + 3\,\hat u^3 + m^2\,\hat u(7\,\hat u -4\,\hat t))\nonumber\\
&&+ d^4\,\hat u( 4\,m^6 + 67\,m^4\,\hat u + \hat u^2(4\,\hat u-\hat t) + m^2\,\hat u( 69\,\hat u + \hat t))]\nonumber\\\nonumber\\
M_{st}&=&\frac{4}{\hat s (\hat t-m^2)}[2\,g_s^4( m^4 -
2\,m^2\,\hat u + \hat t\,\hat u ) - w d^3\,m( m^4 + 4\,m^2\,\hat t + 3\,\hat t^2 ) \,{( -2\,m^2 + \hat t + \hat u ) }^2\nonumber\\
&& + g_s^2 d( 3\,d\,\hat t( m^4 + \hat t\,\hat u + m^2( -3\,\hat
t+
\hat u)) - m( \hat t - \hat u) \,{( -2\,m^2 + \hat t + \hat u)}^2\,w)]\nonumber\\\nonumber\\
M_{su}&=&\frac{4}{\hat s(\hat u-m^2)}[2\,g_s^4( m^4 -2\,m^2\,\hat
t + \hat t\,\hat u) -w d^3\,m\,{( -2\,m^2 + \hat t +
\hat u) }^2( m^4 + 4\,m^2\,\hat u + 3\,\hat u^2)\nonumber\\
 &&+g_s^2 d( 3\,d\,\hat u( m^4 + m^2(\hat t - 3\,\hat u)  + \hat
t\,\hat u) + w m( \hat t - \hat u ) \,{( -2\,m^2 + \hat t + \hat u ) }^2 )]\nonumber\\\nonumber\\
M_{tu}&=&\frac{2}{(\hat t-m^2)(\hat u-m^2)}[4\,g_s^4\,m^2(\hat s
-4\,m^2 ) + d^4\,m^2( 13\,m^6 - 7\,m^4( \hat s - 10\,\hat
u)  + 2\,\hat s\,\hat u(\hat s + \hat u)\nonumber\\
&& - m^2\,\hat u( 39\,\hat s + 35\,\hat u)) + d^2\,g_s^2( -140\,m^6 + \hat s\,\hat u( \hat s +\hat u)  + m^4( 49\,\hat s + 88\,\hat u )\nonumber\\
    &&  -
       m^2( \hat s^2 + 46\,\hat s\,\hat u + 44\,\hat u^2))]\nonumber
\end{eqnarray}

 where $d_1=d_q/2\Lambda$, $d=d_{\tilde g}/4\Lambda$ and
$w=C_W/6\Lambda^2$.

In the special case of $d_1$=$d$=$w$=0 we obtain the following
result for the differential cross section for the gluino
production which agrees with references
\cite{Harrison:1982yi,Dawson:1983fw}:
\begin{eqnarray}
\frac{d\hat{\sigma }(gg \rightarrow
\tilde{g}\tilde{g})}{d\hat{t}}&=&\frac{9\pi\alpha_s^2}{4s^2}
[\frac{2\,\left( m^2 - t \right)\,\left( m^2 - u \right) }{s^2}
-\frac{m^4 - t\,u + m^2\,\left( 3\,t + u \right) }{{\left( m^2 - t
\right) }^2}-
  \frac{m^4 - t\,u + m^2\,\left( t + 3\,u \right) }{{\left( m^2 - u \right)
  }^2}
\nonumber\\
&&+\frac{m^4 - 2\,m^2\,t + t\,u}{s\left(u- m^2\right)} + \frac{m^4
- 2\,m^2\,u + t\,u}{s\left(t- m^2\right)} + \frac{m^2\,\left(
s-4\,m^2 \right) }{\left( m^2 - t \right) \,\left( m^2 - u
\right)}]
\end{eqnarray}
but we should note that the denominators (or numerators) of $s-t$
and $s-u$ terms should be exchanged in Eq.(3.10) of ref
\cite{Dawson:1983fw} and Eq.(3.1) of ref \cite{Harrison:1982yi}.

\newpage

\begin{figure}[h]
  \includegraphics[width=10cm]{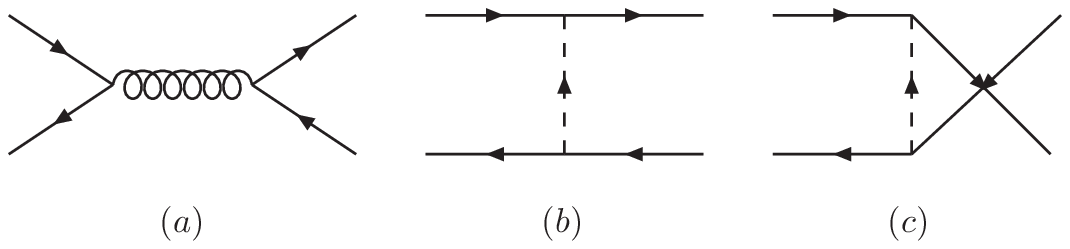}\\
  \caption{$qq\rightarrow \tilde g\tilde g$}
\end{figure}

\begin{figure}[h]
  \includegraphics[width=10cm]{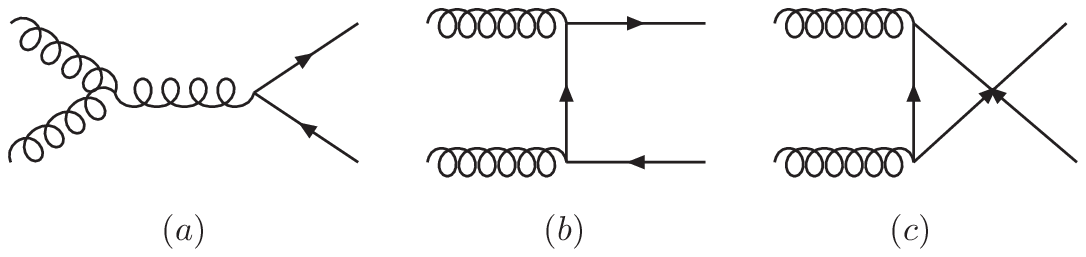}\\
  \caption{$gg\rightarrow \tilde g\tilde g$}
\end{figure}
\newpage
\begin{figure}
  \includegraphics[width=12cm]{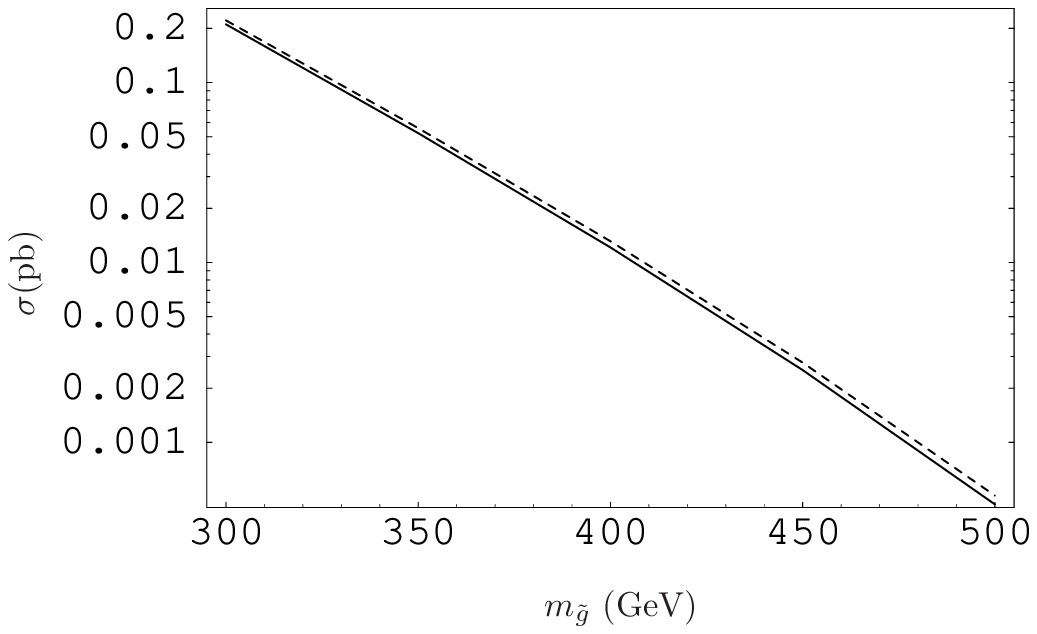}\\
  \caption{The total cross sections for the $qq$ initial states for
the Tevatron with $\Lambda$=1 TeV.}
\end{figure}
\begin{figure}
  \includegraphics[width=12cm]{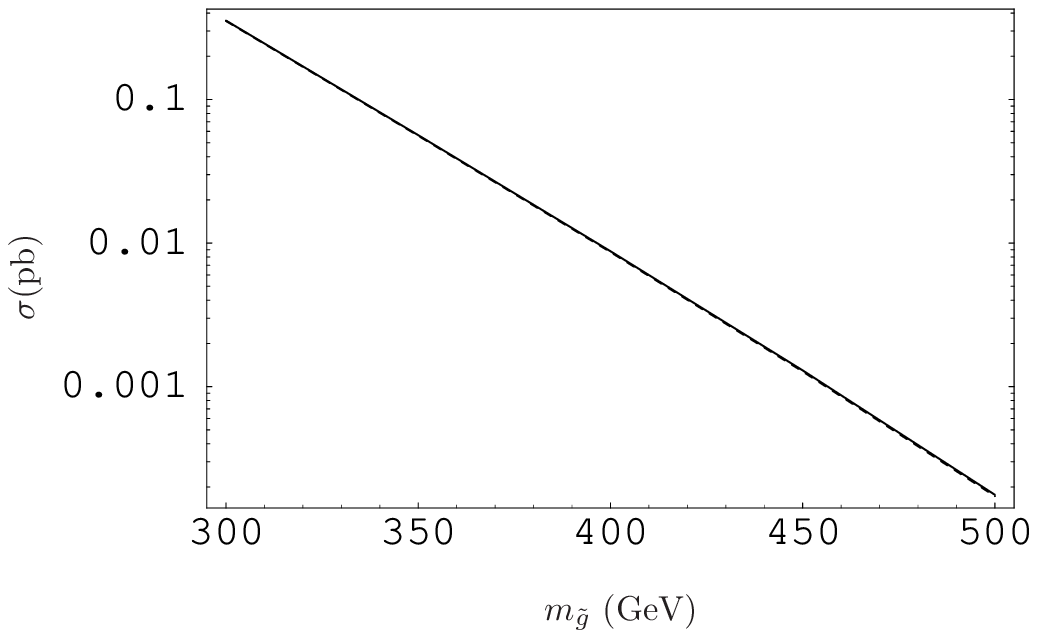}\\
  \caption{The total cross sections for the $gg$ initial states for
the Tevatron with $\Lambda$=1 TeV.}
\end{figure}
\begin{figure}
  \includegraphics[width=12cm]{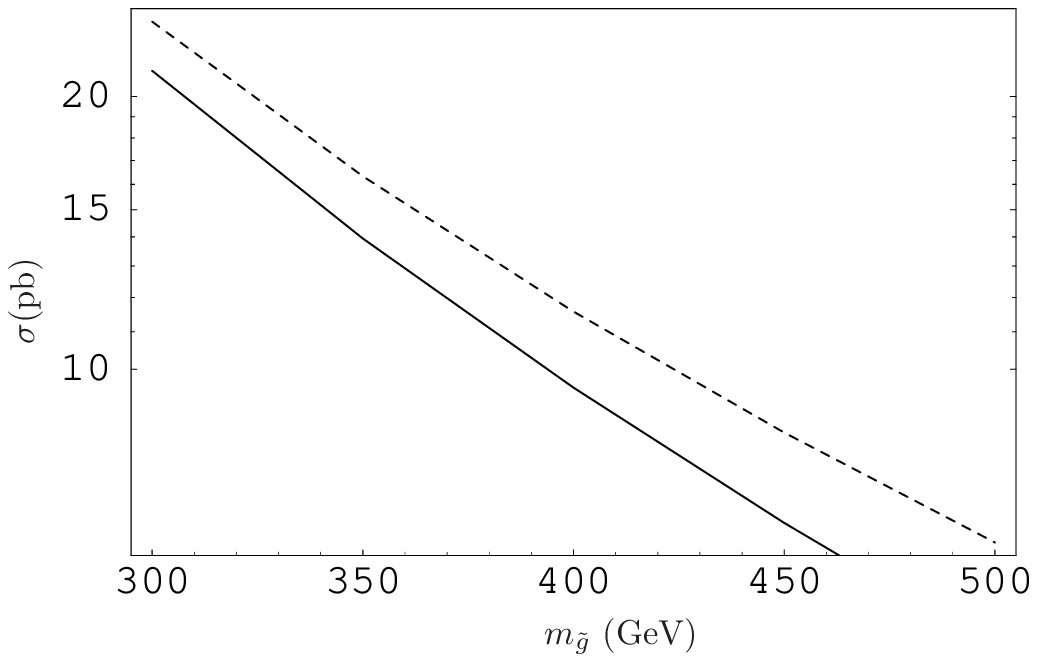}\\
  \caption{The total cross sections for the $qq$ initial states for
the LHC with $\Lambda$=1 TeV.}
\end{figure}
\begin{figure}
  \includegraphics[width=12cm]{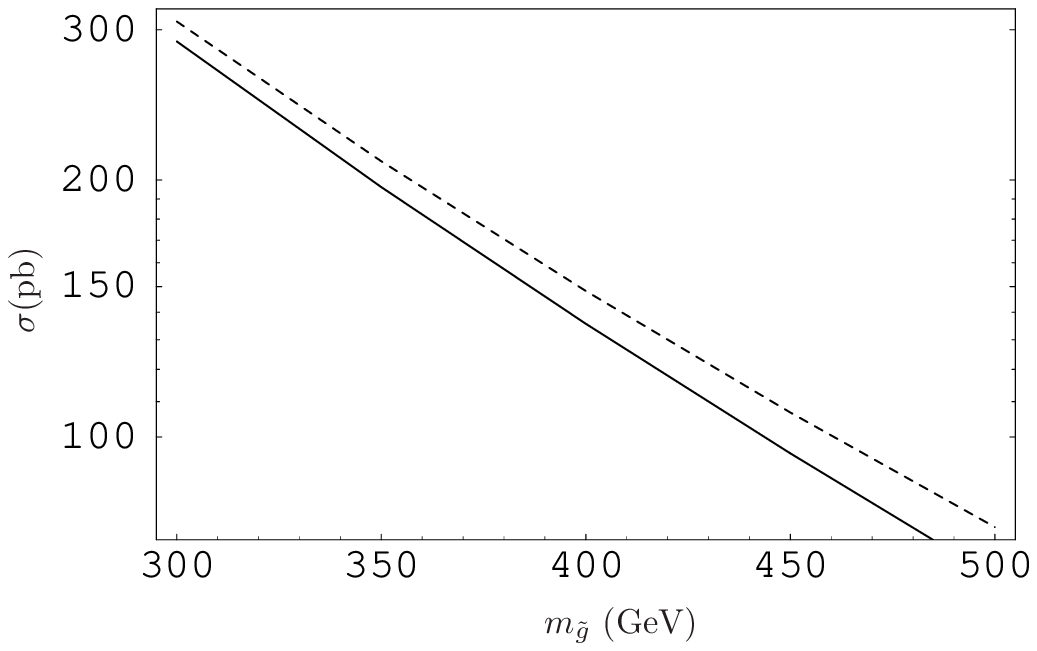}\\
  \caption{The total cross sections for the $gg$ initial states for
the LHC with $\Lambda$=1 TeV.}
\end{figure}


\begin{thebibliography}{99}

\bibitem{HG}
H.~E.~Haber and G.~L.~Kane, Phys. Rep. \textbf{117}, 75 (1985).
\bibitem{HP}
H.~P.~Nilles, Phys. Rep. \textbf{110}, 1 (1984).
\bibitem{LMT}
L.~Girardello and M.~T.~Grisaru, Nucl. Phys. B \textbf{ 194}, 65
(1982).
\bibitem{WRMZ}
W.~Beenakker, R.~Höpker, M.~Spira, P.~M.~Zerwas, Nucl. Phys. B
\textbf{ 492}, 51 (1997).
\bibitem{KL}
G.~L.~Kane, J.~P.~Leveille, Phys. Lett. B \textbf{ 112}, 227
(1982).
\bibitem{SW}
S.~Weinberg, Phys. Rev. Lett. \textbf{63}, 2333 (1989).
\bibitem{DOKM}
D.~A.~Demir, O.~Lebedev, K.~A.~Olive, M.~Pospelov and A.~Ritz,
 Nucl. Phys. B {\bf 680}, 339 (2004).
\bibitem{DMA}
D.~A.~Demir, M.~Pospelov and A.~Ritz, Phys. Rev. D {\bf 67},
015007, (2003).
\bibitem{AP}
  A.~Pilaftsis, Phys. Rev. D {\bf 62}, 016007 (2000).
\bibitem{MRST}
 A.~D.~Martin, R.~G.~Roberts, W.~J.~Stirling and R.~S.~Thorne,
  Eur. Phys. J. C {\bf 4}, 463 (1998).
%
\bibitem{Harrison:1982yi}
 P.~R.~Harrison and C.~H.~Llewellyn Smith,
Nucl.\ Phys.\ B {\bf 213}, 223 (1983)
  [Erratum-ibid.\ B {\bf 223}, 542 (1983)].
\bibitem{Dawson:1983fw}
  S.~Dawson, E.~Eichten and C.~Quigg,
Phys.\ Rev.\ D {\bf 31}, 1581 (1985).
\end{thebibliography}
\end{document}